\documentclass[acmsmall]{acmart}
\usepackage{enumitem}
\usepackage[utf8]{inputenc} 
\usepackage[most]{tcolorbox}
\usepackage[table]{xcolor}
\usepackage{tabularx}
\usepackage{array}
\usepackage{booktabs}
\usepackage{makecell} 
\usepackage{subcaption}
\usepackage{mdframed}
\usepackage{multirow}
\usepackage{caption}
\newmdenv[
  leftmargin=0pt,
  rightmargin=0pt,
  innerleftmargin=1ex,
  innerrightmargin=0pt,
  innertopmargin=0pt,
  innerbottommargin=0pt,
  linewidth=3pt,
  linecolor=gray!40,
  topline=false,
  bottomline=false,
  rightline=false,
  skipabove=2pt,
  skipbelow=2pt
]{mdquote}

\AtBeginDocument{
  }
    
\setcopyright{cc}
\setcctype{by-nc-nd}
\acmDOI{10.1145/3797147}
\acmYear{2026}
\acmJournal{PACMSE}
\acmVolume{3}
\acmNumber{FSE}
\acmArticle{FSE020}
\acmMonth{7}
\acmSubmissionID{fse26maina-p336-p}
\received{2025-09-12}
\received[accepted]{2025-12-22}

\begin{document}

\title[On the Road to Personalized Code Intelligence: Assisting Developers Based on Their In-IDE Behaviors]{On the Road to Personalized Code Intelligence: Portraiting and Assisting Developers Based on Their In-IDE Behaviors}

\author{Yuhong Liu}
\orcid{0009-0009-8747-4885}
\affiliation{%
  \institution{Beihang University}
  \city{Beijing}
  \country{China}
}
\email{katockcricket@buaa.edu.cn}

\author{Yunhe Su}
\orcid{0009-0008-1266-8039}
\affiliation{%
  \institution{Beihang University}
  \city{Beijing}
  \country{China}
}
\email{suyunhe@buaa.edu.cn}

\author{Zhipeng Peng}
\orcid{0009-0005-4258-8521}
\affiliation{%
  \institution{Beihang University}
  \city{Beijing}
  \country{China}
}
\email{hironin@buaa.edu.cn}

\author{Zhiwen Luo}
\orcid{0009-0009-2009-5637}
\affiliation{%
  \institution{Beihang University}
  \city{Beijing}
  \country{China}
}
\email{zhiwenluo@buaa.edu.cn}

\author{Lin Shi}
\authornote{Corresponding author.}
\orcid{0000-0003-1476-7213}
\affiliation{%
  \institution{Beihang University}
  \city{Beijing}
  \country{China}
}
\email{shilin@buaa.edu.cn}

\author{Zhi Jin}
\orcid{0000-0003-1087-226X}
\affiliation{%
  \institution{Wuhan University}
  \city{Wuhan}
  \country{China}
}
\email{zhijin@whu.edu.cn}

\author{Li Zhang}
\orcid{0000-0002-2258-5893}
\affiliation{%
  \institution{Beihang University}
  \city{Beijing}
  \country{China}
}
\email{lily@buaa.edu.cn}

\renewcommand{\shortauthors}{Liu et al.}

\begin{abstract}
With the advent of powerful large language models (LLMs), research in automated software engineering has increasingly focused on leveraging these models to achieve a deeper semantic understanding of code or to engineer sophisticated agent-based processes. The predominant goal of these efforts is to enhance developer productivity through automated assistance. However, this research trajectory has largely overlooked a critical factor: the developers themselves. Programming is a deeply human and individualized activity; developers exhibit significant variation in their coding styles, tool-chain preferences, domain-specific expertise, and problem-solving strategies. Consequently, the current paradigm of one-size-fits-all code intelligence systems struggles to accommodate the unique characteristics and needs of individual developers. To address this gap, we introduce \textbf{VirtualME}, a novel IDE-embedded data infrastructure designed to model the developer by continuously capturing and interpreting their dynamic programming behaviors and preferences.
VirtualME contains three components. (1) Log-level Behavior Extraction: it captures and extracts developers' log-level behaviors (edits, navigations, etc.) from IDE. (2) Task-level Behavior Recognition: it aggregates log-level behaviors into task-level behaviors (``skimming API docs'', ``iterative debugging'', etc.) via a multi-agent pipeline. (3) Developer-persona Measurement: it builds a rule engine to distill a four-dimensional developer persona: Core Technical Foundation, Practical Development Efficiency, Personal Development Norms, and Technical Adaptability. 
On top of VirtualME, we propose a solution for personalized repository-level knowledge Q\&A by integrating the developer persona into a Chain-of-Thought (CoT) guided agent. We evaluated VirtualME by building a multi-repository benchmark with real-world developer trajectories, balancing correctness and personalization. Experimental results show that VirtualME-enhanced answers outperform generic baselines on five dimensions: correctness, cognitive-level fit, technology-stack relevance, behavioral-pattern alignment, and stylistic preference, yielding an average 33.80\% improvement. Our results demonstrate that abundant, continuous developer-behavior data can unlock \textbf{Personalized Code Intelligence}. By integrating this personalized understanding into the code intelligence loop, our approach paves the new way for adaptive and personalized code intelligence.
\end{abstract}




\maketitle

\section{Introduction}
\label{sec:intro}
In recent years, the rapid advancement of large language models (LLMs) has opened new opportunities in automated software engineering~\cite{hou2024large,fan2023large,wang2024software,chen2025deep}.
Current research in this field primarily focuses on two directions.
On the one hand, LLMs are employed to achieve a deep semantic understanding of code~\cite{mathews2024test,nam2024using,shi2025natural,mu2024clarifygpt}, which enables tasks such as accurate bug detection~\cite{guan2025crossprobe, sagodi2024reality}, intelligent code optimization~\cite{gao2024search, rahman2025marco,wu2024ismell}, and reliable code completion~\cite{wang2024llms, eghbali2024hallucinator}, etc.
By reducing human errors and improving the efficiency of routine programming tasks, this capability accelerates the development cycle and enhances software robustness~\cite{yang2025advancing}.
On the other hand, LLMs are increasingly integrated into agent-based workflows~\cite{yang2024swe,he2025llm,ronanki2025facilitating}, where they can coordinate complex development processes, automate issue resolution, and support collaborative software maintenance. Such workflow automation not only lowers the overhead of project management~\cite{bouzenia2025you} but also fosters more scalable and adaptive engineering practices, ultimately leading to higher developer productivity and improved software quality~\cite{liu2025rtadev}.

{However, a critical oversight persists in existing automated software engineering research: it has largely sidelined the end-users of these technologies—the developers themselves. Code intelligence systems, by design, are built to support developers; yet their utility hinges entirely on whether the assistance they deliver is validated and adopted by the very individuals they aim to serve~\cite{abrahao2025software,sergeyuk2025using}. Importantly, developers exhibit substantial heterogeneity across multiple dimensions: coding style variations, toolchain preference discrepancies, varying levels of domain-specific expertise, and divergent problem-solving methodologies. Failing to account for these individual differences directly compromises both the effectiveness of code intelligence techniques and their likelihood of widespread adoption~\cite{fakhrhosseini2024user}. This limitation is particularly pronounced in the domain of repository-level Q\&A—a task that encompasses inquiries ranging from code debugging and architectural design to version control and collaborative problem-solving—where personalized information exerts a more significant influence than any other factor.}
% \lin{add citations at somewhere}

Here we present a typical scenario to illustrate the limitations of code intelligence systems in repository-level question-answering when lacking personalized context.
Two developers——a senior architect and a junior engineer——pose the same question to a code intelligence system (e.g., GitHub Copilot): ``Please describe the authorization module.''
Because the system lacks an understanding of developers' individualized characteristics, it typically produces a unified response, starting with basic concepts (such as role-based access control, RBAC) and subsequently detailing the specific classes and APIs used in the implementation.
For the senior architect, this lengthy response is suboptimal, as most of the content is already within their domain knowledge, requiring them to manually filter information to extract new implementation details.
In contrast, for the junior engineer, due to a lack of domain knowledge (e.g., the purpose of RBAC), the same response is difficult to comprehend, and the junior engineer must engage in multiple rounds of clarifying interactions to fully understand the system's unified response.

The above gaps prompt us to inquire: \textbf{\textit{how can code intelligence systems effectively account for individual developer differences and adapt their solutions to each developer's unique characteristics?}}
In fact, a developer's dynamic behaviors within the Integrated Development Environment (IDE)—their patterns of code authoring, navigation pathways, and debugging trajectories—implicitly encode a rich representation of their personal competencies and habits. If these behavioral signals can be systematically captured and interpreted, they could evolve the paradigm of code intelligence from ``one-size-fits-all'' to ``one-size-fits-one''.

Building on this insight, we propose \textbf{VirtualME}, a novel IDE-embedded data infrastructure designed to construct a dynamic model of the developer by continuously capturing and interpreting their programming behaviors.
First, to ensure that all our analyses of the developer are founded upon a precise and verifiable digital footprint, VirtualME captures the developer's raw interactions within the IDE as \textbf{Log-level Behaviors} (LBs), such as edits, navigations, and terminal executions. This data constitutes the most objective and unbiased ground truth of a developer's activity.
However, log-level behaviors are too fine-grained and lack deep semantic meaning. To bridge the gap between atomic actions and developers' intent, we introduce an intermediate layer of \textbf{Task-level Behaviors} (TBs). We employ a multi-agent pipeline to aggregate sequences of LBs into meaningful tasks (e.g., ``attempting to resolve an exception''). Within this pipeline, a clustering algorithm effectively identifies temporally and contextually related behavior groups, while an LLM agent excels at summarizing these groups into high-level task descriptions—a highly flexible form of generalization that is difficult to achieve with traditional classification approaches.
Finally, based on various cognitive theories of programmer behavior that define and measure developer characteristics and proficiency, we design a lightweight rule engine. This engine distills the collected task-level behaviors into a comprehensive, four-dimensional developer persona, encompassing their: Core Technical Foundation, Practical Development Efficiency, Personal Development Norms, and Technical Adaptability. This rule-based approach ensures our computation process is both transparent and verifiable.

To validate the practical value of our approach, we implement a repository-level Q\&A module built upon VirtualME, which leverages the developer persona aiming to generate answers that are precisely aligned with the user's expertise and preferences. To quantify the improvements of this approach over baselines, we construct a benchmark based on four repositories and evaluate along five dimensions, jointly measuring correctness and personalization. Furthermore, we collect four weeks of IDE behavioral traces from 10 developers, providing each agent with the corresponding behavioral data infrastructure. Comparative experiments demonstrate that our approach outperforms the baseline by an average of 33.80\%, highlighting the utility of IDE behavioral data in advancing \emph{Personalized Code Intelligence}.

This paper makes the following contributions:
\begin{itemize}[leftmargin=1em]
    \item \textbf{IDE infrastructure for understanding developer behaviors.} We design a system that models developer behavior at two complementary log-level behaviors and task-level behaviors—and, building on these representations, distills a comprehensive developer persona to support \emph{Personalized Code Intelligence}. 
    \item \textbf{Public dataset of developer IDE behavioral traces.} We construct and release a dataset of IDE-based developer behavioral traces, enabling systematic analysis of development practices.
    \item \textbf{Solution for Personalized repository Q\&A.} We propose a solution for personalized developer Q\&A, together with a new benchmark and a public dataset that jointly evaluates correctness and personalization, providing a standardized basis for future research.
\end{itemize}

The remainder of the paper is organized as follows. Section~\ref{sec:theory} defines \emph{Personalized Code Intelligence}, proposes a representation model for developer persona based on the established theoretical framework, and presents a cognitive-theory-based model of in-IDE developer behavior. Section~\ref{sec:approach} presents VirtualME, which extracts log-level behaviors, aggregates them into task-level behaviors, derives rule-based developer personalities, and applies them to personalized repository Q\&A. Sections~\ref{sec:exp} and ~\ref{sec:results} describe the Q\&A benchmark and user study for quantitative evaluation. Section~\ref{sec:discussion} discusses future directions, Section~\ref{sec:rw} reviews related work, and Section ~\ref{sec:conclusion} summarizes this work.

\section{Preliminary Study}
\label{sec:theory}
To ground our approach in a clear and formal basis, this section establishes the foundational definitions upon which our work is built. We first formalize the concept of \emph{Personalized Code Intelligence}, defining the role of a developer persona. 
We then synthesize established theories of programmer competence to propose a four-dimensional representation of a developer’s persona.
Finally, we introduce the \emph{Log-level Behavior} model, our schema for capturing the fine-grained, in-situ data from IDE activities that are required to instantiate and continuously refine this persona. These definitions collectively form the theoretical bedrock for our subsequent analysis of IDE behaviors and their application to \emph{Personalized Code Intelligence}. 

\subsection{Definition of Personalized Code Intelligence}
\label{subsec: defPers}
We ground our investigation in a precise, formal definition of \emph{Personalized Code Intelligence} that clarifies how developer-centric information augments conventional IDE-based assistance.

\paragraph{Conventional IDE Intelligence}
Existing code intelligence systems can be abstracted as a mapping:
\begin{equation}
  \label{eq:conventional}
  \mathcal{O} = \pi(\mathcal{C}, \mathcal{I})
\end{equation}
where:
\begin{itemize}[leftmargin=1em]
  \item $\pi$ represents the \emph{code intelligence system} implemented by the underlying large model;
  \item $\mathcal{O}$ denotes the system’s final output (e.g., generated code, patch, or explanation);
  \item $\mathcal{C}$ is the \emph{context}, comprising repository-level artifacts (files, ASTs, import graphs) and any user-supplied situational context;
  \item $\mathcal{I}$ is the \emph{instruction}, encapsulating prompts, user queries, or task descriptions.
\end{itemize}

\paragraph{Personalized IDE Intelligence}
We extend Eq.~\eqref{eq:conventional} by injecting a developer-specific context $\mathcal{P}$—the \emph{Persona} derived in Section~\ref{subsec:persona}—to yield:
\begin{equation}
  \label{eq:personalized}
  \mathcal{O}_{\text{pers}} = \pi(\mathcal{C}, \mathcal{P}, \mathcal{I})
\end{equation}
The additional term $\mathcal{P}$ includes personalized information extracted from the developer's IDE behaviors, such as their knowledge and technical domain, work efficiency, habits, and learning adaptability. By conditioning the agent on $\mathcal{P}$, the system can: (1) anticipate latent developer needs; (2) align suggestions with individual work styles; (3) prioritize information that has historically accelerated the developer’s workflow; and (4) pay special attention to tasks where the developer has historically encountered difficulties. Consequently, this formula formalizes the transition from context-agnostic assistance to \emph{Personalized Code Intelligence}.

\subsection{Representation of Developer Persona}

\label{subsec: RepPC}
 
In empirical software engineering research, mapping programmer behavior to individual characteristics has been a central theme for process improvement and personnel assessment~\cite{CRUZ201594}. We review and synthesize prevailing descriptive frameworks of developer cognitive behavior and competence to provide a systematic theoretical basis for the subsequent modeling of developer behavior and the quantification of developer persona metrics in later sections.

Our approach is grounded in a synthesis of established theoretical frameworks that connect developer actions to competence and productivity. The Personal Software Process (PSP)~\cite{humphrey2005psp} establishes a data-driven foundation, quantifying performance through metrics like productivity and defect rates based on time and size tracking. Complementing this, the IEEE Software Engineering Competency Model (SWECOM)~\cite{swecom2014} provides a cognitive perspective, defining the reasoning and problem-solving skills that underpin complex development tasks. To capture the broader developer experience, we draw from the SPACE framework~\cite{forsgren2021space}, which models productivity across dimensions like Activity and Efficiency, and DevEx framework~\cite{noda2023devex}, which explains the impact of cognitive load and flow state on performance. Finally, to ground our understanding of learning behaviors, the Introductory Programming Self-Efficacy Scale (IPSES)~\cite{introProgSEScale2020} offers psychological constructs for assessing how developers approach and learn from unfamiliar code.

The existing theories form a theoretical puzzle for multi-dimensional persona measurement: PSP supplies a quantitative performance baseline; SWECOM enriches the cognitive-ability view; SPACE and DevEx link behavior to experience; IPSES furnishes psychometric evidence for learning behavior. This integrated perspective underpins our research, which originates from programmer IDE behavior, and enables the mapping from discrete IDE events to an interpretable developer persona. Combined with information provided by developer behavior within the IDE, we orthogonally classify all observable developer characteristics within the IDE into four dimensions: \textbf{Core Technical Foundation (CTF)}, \textbf{Practical Development Efficiency (PDE)}, \textbf{Personal Development Norms (PDN)} and \textbf{Technical Adaptability (TA)}. 
Each dimension comprises multiple metrics:
$$
\text{Persona} = (\text{CTF}, \text{PDE}, \text{PDN}, \text{TA})
$$
where:
\begin{itemize}[leftmargin=1em]
    \item \textbf{CTF} (Core Technical Foundation) focuses on the developer's programming domain, frequently used programming languages, and associated libraries/frameworks. This forms the foundational skill set supporting their development work.
    \item \textbf{PDE} (Practical Development Efficiency) centers on code productivity, task execution success rates, and time-to-fix-failures. This reflects the practical efficiency and quality of their problem-solving.
    \item \textbf{PDN} (Personal Development Norms) includes comment writing frequency, active hours, and preferences for specific plugins and IDE shortcut features. This captures their personal development customs and tool usage tendencies.
    \item \textbf{TA} (Technical Adaptability) involves the cognitive load during navigation of unfamiliar repositories and the frequency of adopting new libraries, frameworks, and languages. This measures their ability to cope with unknown technologies and acquire new skills.
\end{itemize}

\subsection{Log-level Behavior Model of In-IDE Developer Activity}
\label{subsec:lb-model}
Most prior work on developer personas relies on post-hoc questionnaires or repository commits/snapshots~\cite{kuutila2018using}~\cite{kuutila2021individual}.
These tools provide only coarse, low-frequency signals and cannot capture the moment-to-moment behavioral and cognitive states that arise during live development.  
Consequently, existing personas lack the temporal granularity and situational fidelity required to drive in-situ personalized assistance.

To obtain the fine-grained, continuous stream of behaviors needed to populate $\mathcal{P}$ in Eq.~\eqref{eq:personalized}, we establish a precise, machine-interpretable model of developer activity inside the IDE.  
We model every developer–artifact interaction as a \textbf{Log-level Behavior (LB)}—a timestamped triple (Action, Object, Context)—that preserves the complete time series while remaining amenable to automated analysis.  
This resolution data foundation yields both data consistency (uniform schema across tools) and interpretability (action-level semantics), enabling further subsequent mapping from raw events in IDE to the higher-level personal characteristics in $\mathcal{P}$.

\begin{equation}
\text{Behavior} = (\text{Timestamp}, \text{Action}, \text{Object}, \text{Context})
\end{equation}

\noindent
\textbf{Timestamp} is the exact time of the behavior; \textbf{Action} is the specific operation, such as ``add code'' or ``execute terminal command''; \textbf{Object} is the target of the action, e.g., a function or a file; \textbf{Context} supplies the surrounding circumstances, such as the content of the code change or the terminal command and its output.

With this model, we can comprehensively describe a programmer’s log-level IDE behavior and, on that basis, extract deeper, latent features.

\section{Approach}
\label{sec:approach}

\begin{figure}
  \centering
  \includegraphics[width=\linewidth]{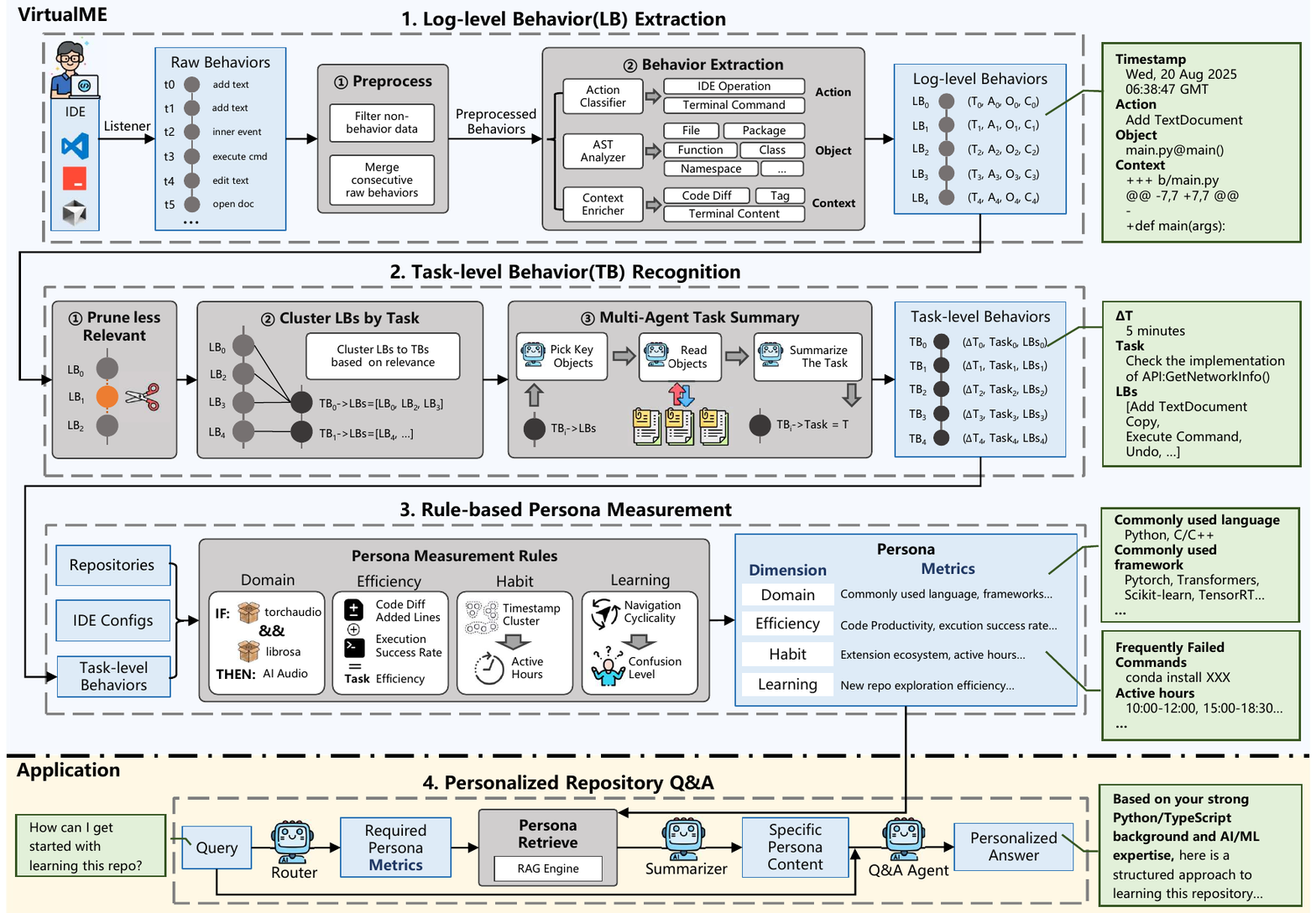}
  \caption{The architecture of VirtualME and the solution of personalized Repository Q\&A.}
  \label{fig:vm}
\end{figure}

Based on our definition of \emph{Personalized Code Intelligence}, we propose \textbf{VirtualME}, a data infrastructure within the IDE to systematically analyze developer behavior data. As illustrated in Figure~\ref{fig:vm}, VirtualME consists of three main modules. (1) \textbf{Log-level Behavior (LB) Extraction} module extracts a developer's fine-grained actions as Log-level Behaviors in real-time. (2) \textbf{Task-level Behavior (TB) Recognition} module employs a multi-agent pipeline to recognize the specific development tasks being performed from these LBs. (3) After a period of collection and recognition, \textbf{Rule-based Persona Measurement} module computes a detailed developer persona. Additionally, we design (4) a Repository Q\&A approach that guides an agent with a Chain-of-Thought (CoT)~\cite{wei2022chain} process, leveraging the VirtualME data infrastructure to generate responses aligned with the developer's preferences.

\subsection{Log-level Behavior Extraction}
\label{subsec:lb}

To achieve comprehensive monitoring of a developer's activities within the IDE and support in-depth analysis, we first collect and extract a developer's Log-level Behaviors (LBs) from the IDE, in accordance with the definition provided in Section~\ref{subsec:lb-model}. We utilize the VSCode API to listen for all behaviors occurring in the IDE. These behaviors, triggered by the developer, the IDE itself, or an LLM agent, are collected as \textit{Raw Behaviors}, containing details such as the behavior name, associated files and line numbers, and behavior outputs. These \textit{Raw Behaviors} are then processed by a dedicated \textit{Preprocess} unit, which filters out actions not initiated by the programmer and performs an initial consolidation of consecutive editing actions.

The preprocessed behaviors are then passed to the \textit{Behavior Extraction} unit, which transforms them into programmer behaviors represented as quadruples (Timestamp, Action, Object, Context), as defined earlier (see Section~\ref{subsec:lb-model}).

\begin{itemize}[leftmargin=1em]
    \item \textbf{Action:} We employ a rule-based classifier to categorize the preprocessed behaviors into developer actions, including \textit{IDE Operations} (e.g., adding or deleting text) and \textit{Terminal Commands} (e.g., executing a command). This constitutes the \textit{Action} in our LB model.
    \item \textbf{Object:} Using the Abstract Syntax Tree (AST) analyzer from the corresponding language server, we resolve the raw files and line numbers into specific code artifacts, which represent the \textit{Object} in our LB model, such as files, classes, or functions.
    \item \textbf{Context:} We employ a rule-based engine named \textit{Context Enricher} to tag the input and output of terminal events (e.g., command type, domain, success status) and record the code changes of editing behaviors. This constitutes the \textit{Context} in our LB model.
\end{itemize}

Through this process, we successfully capture a comprehensive and fine-grained stream of the developer's LBs from the IDE. This real-time, detailed record of all granular actions lays the foundation for subsequent, more profound analysis.

\subsection{Task-level Behavior Recognition}
\label{subsec:tb}

In the previous module, we collect a developer's Log-level Behaviors, which are atomic, making it hard to infer the developer's deeper intentions or the full scope of their programming task from a single behavior. It is therefore necessary to adopt a holistic perspective to understand a developer's Task. We innovatively use a combination of clustering algorithms and LLM agents to aggregate and identify LBs as higher-level activities, termed \textbf{Task-level Behavior (TB)}. A TB offers a more concise summary of a developer's activities over a period and is composed of several contiguous LBs.
$$
TB = (\Delta T, \text{Task}, LBs), \quad \text{where } LBs = [LB_0, LB_1, LB_2, \dots]
$$
Here, $\Delta T$ is the duration of the TB, calculated from the timestamps of the first and last LBs. \textit{Task} is a natural language description of the development task performed during this period, and \textit{LBs} is the array of the LB themselves.

The recognition of a TB begins with pruning and clustering a real-time stream of LBs. We first calculate the relatedness score, $R$, between LBs across four dimensions: 
\textit{Time (T)}, \textit{Action (A)}, \textit{Object (O)}, and \textit{Context (C)}. For example, the formula for calculating the relatedness score between $LB_x$ and $LB_y$ is as follows:
$$
R_{xy} = a \cdot R(T_x, T_y) + b \cdot R(A_x, A_y) + c \cdot R(O_x, O_y)  + d \cdot R(C_x, C_y)
$$
\begin{itemize}[leftmargin=1em]
    \item The temporal similarity $ R(T_x, T_y)$ is based on the time interval; behaviors with smaller time gaps are more likely to be clustered. 
    \item The object similarity $R(O_x, O_y)$ is determined by the distance between two objects in the AST; LBs on closely related objects are more likely to be clustered. 
    \item To compute the semantic similarity of actions and context ($R(A_x, A_y)$ and $R(C_x, C_y)$), we use the ~\href{https://huggingface.co/sentence-transformers/all-MiniLM-L6-v2}{all-MiniLM-L6-v2} embedding model with cosine similarity, which is a lightweight, general-purpose semantic embedding model provided by Sentence Transformers, selected for its balance between efficiency and accuracy. Since actions and context consist of code snippets and program outputs—which differ from natural language in structure and semantics—off-the-shelf models yield suboptimal results. Therefore, we fine-tuned the model on 10k+ manually annotated log-level behaviors from nine internal developers. This fine-tuned model achieves an accuracy of over 89\% on internal validation set, enabling stable identification of semantic relationships between actions and contexts. 
    \item The adjustable weights ($a$=0.35, $b$=0.3, $c$=0.25, $d$=0.1) were empirically tuned prior to the user study using the manually annotated log-level behaviors mentioned above, and kept fixed thereafter for consistency and reproducibility.
\end{itemize}
The \textit{pruning} step first removes LBs with low similarity to their neighbors, which typically represent noise unrelated to the primary task. Subsequently, the remaining LBs undergo density-based clustering using the Density-Based Spatial Clustering of Applications with Noise (DBSCAN)~\cite{ester1996density} algorithm—chosen for its ability to handle a variable number of clusters. Related LBs are clustered into an array, forming the $LBs$ attribute of a new TB, signifying that they all serve the same task. To support multi-task recognition in real-world development, this also allows non-adjacent LBs belonging to the same task to be merged.

To populate the \textit{Task} field by summarizing the developer's high-level activity from the LBs, we design a multi-agent stream:
\begin{enumerate}[leftmargin=2em]
    \item The first agent selects \textit{Key Objects} from the LBs that it deems critical for task recognition (e.g., a project's configuration file during environment setup).
    \item The second agent retrieves the corresponding source code snippets from the repository, using file-reading permissions.
    \item The third agent synthesizes a summary from the retrieved code snippets, incorporating additional context from the LBs such as terminal logs.
\end{enumerate}
For instance, for a sequence of LBs involving continuous code editing, the multi-agent stream might read the affected files and code diffs and summarize the Task as ``Implemented a new API feature.'' For a sequence involving code edits and terminal executions, the multi-agent stream could analyze the code, execution logs, and summarize the \textit{Task} as ``Attempted to fix exception X thrown in code segment A.''

\subsection{Rule-based Persona Measurement}
\label{subsec:persona}

TBs summarize \emph{what} a developer did during a short interval; they do not yet capture \emph{who} the developer is. Therefore, we derive the developer's persona—an interpretable, theory-grounded profile from the accumulated TB stream.  
Building on the theory foundation reviewed in Section~\ref{sec:theory}, we distill each developer’s IDE telemetry into five orthogonal dimensions: \textbf{Core Technical Foundation (CTF)}, \textbf{Practical Development Efficiency (PDE)}, \textbf{Personal Development Norms (PDN)} and \textbf{Technical Adaptability (TA)} (see Section~\ref{subsec: defPers}).During design, we consulted three senior SE researchers familiar with the theoretical frameworks (PSP, SWECOM, etc.) to align and refine the metrics, ensuring all metrics are traceable to theory and unambiguously extractable from IDE logs.
The derivation is rule-based (see Table~\ref{tab:personality-metrics}). The system updates the developer's persona periodically (empirically set to daily) based on newly collected TBs.
 
\begin{table*}
  \caption{Metrics for Developer Persona Measurement (Part).}
  \label{tab:personality-metrics}
  \renewcommand{\arraystretch}{1.2}
  \small
  \setlength{\tabcolsep}{3.5pt}
  \begin{tabular}{@{}p{0.6cm} p{1.8cm} p{8.0cm} p{2.4cm}@{}}
    \toprule
    Dim & Metric & Rules & Example\\
    \midrule
    CTF & Frequent Libraries &
    $\mathcal{F} = \text{Classify}(\mathcal{P}_{\text{ext}})$, 
    where $\mathcal{F}$ is the set of frameworks and $\mathcal{P}_{\text{ext}}$ is the set of external packag-es. &
    PyTorch \\

    \cmidrule(lr){2-4}
    & Knowledge Domain &
    $\text{IF } (f_1 \in \mathcal{F} \land f_2 \in \mathcal{F}) \text{ THEN } \mathcal{D}_1$, where $f_i$ is a framework and $\mathcal{D}_1$ is a knowledge domain. &
    LLM Technology \\

    \hline
    PDE & Code \newline Productivity &
    $\text{Prod}_{L} = \frac{\Delta L_{\text{added}}}{\Delta t}$, 
    where $\Delta L_{\text{added}}$ is the lines added from git diffs over time period $\Delta t$. &
    C++: avg 50 lines per hour \\

    \cmidrule(lr){2-4}
    & Terminal Task Success Rate &
    $\text{SR}_{T} = \frac{N_{T, \text{succ}}}{N_{T, \text{total}}}$, 
    where $N$ is the command count for task type $T$. &
    Python package install: 76\% avg. success \\

    \hline
    PDN & Active Hours &
    $\mathcal{C}_t = \text{Cluster}(\{t_i | t_i \in \text{LBs}\})$, 
    where $\mathcal{C}_t$ are the resulting time clusters from LB timestamps $t_i$. &
    15:00--20:00, Mon--Fri \\

    \cmidrule(lr){2-4}
    & Commenting Frequency &
    $\text{CR}_{L} = \frac{\Delta L_{\text{comment}}}{\Delta L_{\text{code}}}$, where CR is the Commenting Ratio for language $L$. &
    6 comment lines per 10 LoC for Python \\
    
    \hline
    TA & Repo \newline Exploration Efficiency &
    Compute navigation cyclicality: 
    $C = 1 - \frac{P_c}{N}$, where $P_c$ is the position of the current file in history and $N$ is the total files visited. Lower values indicate more divergent navigation. &
    Cyclicality 0.67 in repo X (last 20 min) \\

    \cmidrule(lr){2-4}
    & New Library Adoption Rate &
    $\text{Rate}_{\text{lib}} = \frac{|\mathcal{L}_{t_1} \setminus \mathcal{L}_{t_0}|}{t_1 - t_0}$, where $\mathcal{L}$ is the set of used libraries at time $t$. &
    One new library per week \\
    \bottomrule
  \end{tabular}
\end{table*}

\subsection{Personalized Repository Q\&A}

To validate the rationale and practical value of our proposed developer behavior modeling, recognition, and persona measurement, we design a \textbf{Personalized Repository Q\&A} approach built upon the four-dimensional developer persona provided by VirtualME.
This approach employs Chain-of-Thought (CoT) prompting to guide an agent in leveraging this profile to generate answers that align with the developer's preferences.

In this scenario, developers typically ask a code intelligence system (e.g., GitHub Copilot) knowledge-based questions about a repository, ranging from global queries like ``Analyze the overall project structure'' or ``How do I build this project?'' to local ones like ``What does this function do?'' or ``Analyze the call graph of this API.'' Mainstream approaches have the system answer solely based on static information within the codebase, disregarding the questioner's technical proficiency and comprehension level. This results in standardized answers that offer inconsistent value to experts and novices alike; the former may skip large sections of known information, while the latter may need to repeatedly ask for clarification on concepts they do not understand.

In our approach, the \textit{Router} agent first infers the \textit{Persona Types} likely required to answer the user's \textit{Query}. For example, if a user asks, ``How do I run this App project?'', the agent needs to know which projects the user has run before and what kinds of problems the user has encountered.
$$
M_p = \theta(Q, P)
$$
Here, $\theta$ is the \textit{Router} agent, $Q$ is the user's \textit{Query}, $P$ is an instructional prompt outlining the structure of the developer persona to inform the agent of the available information dimensions, and $M_p$ are the \textit{Required Persona Metrics} (e.g., ``User's tech stack'', ``Proficiency in Swift'').

Next, a simple RAG engine queries the \textit{Persona} to retrieve the specific content relevant to the \textit{Required Persona Metrics}, yielding the user information needed to answer the question.
$$
P_r = \text{Retrieve}(M_p, \text{Persona})
$$
Here, \textit{Persona} is the complete user profile, $M_p$ are the \textit{Required Persona Metrics}, and $P_r$ is the \textit{Specific Persona Content} needed (e.g., ``User's tech stack: Android application development.'', ``Proficiency in Swift: cumulative 523 lines of code'').

Finally, $P_r$ is fed into the \textit{Q\&A Agent} along with the original Query, guided by an instruction to generate an answer tailored to the user's preferences.
$$
O = \theta(P_r, I, Q, W)
$$
Here, $\theta$ is the \textit{Q\&A Agent}, $O$ is the final output, $W$ is the Workspace (i.e., repository knowledge), and $I$ is the guiding instruction.

Through this approach, the developer behavior data foundation we constructed can be fully utilized in a repository Q\&A scenario. In the following section, we describe a series of experiments to carefully evaluate the improvements our approach offers over baseline approaches.
\section{Experimental Design}
\label{sec:exp}
Our evaluation investigates the following research questions:

\begin{itemize}[leftmargin=1em]
    \item \textbf{RQ1} Does the developer persona generated by VirtualME align with the developers' self-assessed ground truth, and how quickly does it converge?
    
    \item \textbf{RQ2} Does the Personalized Repository Q\&A, built upon the VirtualME data foundation, enhance the user's personalized experience without compromising correctness?
    
    \item \textbf{RQ3} When a developer’s proficiency evolves from complete beginner to preliminary mastery over the course of a week, how do the personalized answers dynamically adapt?
\end{itemize}

\subsection {Participant Recruitment \& Overall Observation Procedure}
\label{subsec:participant-recruit}
We conducted human experiments to validate VirtualME. To ensure the reliability and validity, we applied strict recruitment criteria. Participants were required to (1) demonstrate proven ability to navigate and modify codebases exceeding 20k LOC (Lines of Code), (2) represent diverse backgrounds in languages, domains, and experience levels. Since VirtualME relies on stable behavioral data, participants also (3) committed to 4 weeks of continuous usage.

Based on these criteria, we recruited 10 developers ranging from junior to senior levels, covering domains such as backend systems, client applications, and AI development. All participants gave informed consent provided that their data were anonymized. Each of them was compensated with approximately {\$40} for their involvement. Participant details are summarized in Table~\ref{tab:participants}.

\begin{table*}[ht]
  \caption{Participant Demographics.}
  \label{tab:participants}
  \small
  \newcommand{\tabincell}[2]{\begin{tabular}[c]{@{}#1@{}}#2\end{tabular}}
  \begin{tabular}{@{}llllll@{}}
    \toprule
    ID & Domain & Coding Language & Task during experiment & Exp./Year & Degree   \\
    \midrule
    P0 & SE     & JS, TS          & Cross-platform Desktop Apps Development & 3 & Bachelor \\
    P1 & AI     & Python          & Voice Cloning                             & 4 & Bachelor \\
    P2 & AI     & C++             & High-performance Operators Development  & 3 & Bachelor \\
    P3 & AI, SE & JS, TS, Python  & IDE Extension Development                 & 5 & Master   \\
    P4 & SE     & JS, HTML        & Web Frontend Development                  & 4 & Bachelor \\
    P5 & SE     & Java            & Web Backend Development                   & 3 & Bachelor \\
    P6 & SE     & C\#, Python     & Windows Desktop Apps Development          & 1 & Bachelor \\
    P7 & SE     & JS              & IDE Extension Development                 & 4 & Bachelor \\
    P8 & AI     & Python          & Data Batch Processing                     & 5 & Master   \\
    P9 & SE     & PHP             & Web Frontend Development                  & 5 & Bachelor \\
    \bottomrule
  \end{tabular}
\end{table*}

Over the course of 4 weeks, participants carried out their routine development tasks, while the VirtualME plugin unobtrusively collected behavioral data in the background:
\begin{itemize}[leftmargin=1em]
    \item Engineers used the plugin as part of their daily work.
    \item Students used the plugin while completing their course project (large, team-based engineering assignments).
\end{itemize}
Guided by the requirements of our 3 RQs, we designed the following procedure to ensure comprehensive and systematic data collection.

\begin{itemize}[leftmargin=1em]
    \item \textbf{Week 0:} Participants installed the VirtualME plugin, which began collecting their behavioral data in the background.
    \item \textbf{Weeks 1--4:} The plugin continuously monitored participant activities. Every Friday evening, developers reviewed their generated persona and completed a questionnaire to provide feedback on the accuracy of each metric (\textbf{RQ1}).
    \item \textbf{Week 4:} We administered the repository Q\&A test and performed evaluation on both accuracy and personalization.(\textbf{RQ2}).
    \item \textbf{End of Week 4:} Participants could voluntarily enroll in a longitudinal case study to investigate the dynamic evolution of their persona as their behaviors changed (\textbf{RQ3}).
\end{itemize}

\subsection{Dataset for Personalized Code Repository Q\&A}
To rigorously evaluate both the correctness and personalization of our Personalized Repository Q\&A solution, we constructed a dedicated benchmark which includes development questions drawn from multiple repositories and an associated scoring scheme for each question.

\noindent\textbf{Code Repository Questions.} To create a benchmark that reflects realistic and diverse repository Q\&A scenarios, we selected four open-source repositories covering different languages and domains. None of the participants had prior exposure to the chosen projects. These repositories are summarized in Table~\ref{tab:repositories}. To ensure comprehensive coverage of question types, we constructed seven questions for each repository, for a total of 28 questions in the dataset. Each set of seven questions comprised: 
\begin{enumerate}[leftmargin=2em]
    \item \emph{Three global questions} focused on project-level understanding such as architecture, build workflow, and runtime configuration; 
    \item \emph{Four local questions} focused on detailed code-level comprehension, targeting specific files, classes, or functions.
\end{enumerate}
All questions were derived from the repositories’ source contents and distilled from typical inquiries appearing in their issue discussions. Representative examples are shown below.
\vspace{4pt}

\begin{mdframed}[
  linecolor=gray!60,
  linewidth=0.6pt,
  roundcorner=10pt,
  backgroundcolor=gray!5,
  innertopmargin=6pt,
  innerbottommargin=6pt,
  innerleftmargin=10pt,
  innerrightmargin=8pt,
]
\noindent \texttt{Global Question:} What is the overall workflow of this server? \\
\texttt{Local Question:} What is the role of Class A in handling streaming requests?
\end{mdframed}
\vspace{4pt}

\begin{table*}
  \caption{Summary of Selected Repositories for Personalized Code Q\&A Benchmark.}
  \label{tab:repositories}
  \small 
  \begin{tabular}{p{0.05\linewidth}p{0.1\linewidth}p{0.15\linewidth}p{0.5\linewidth}p{0.05\linewidth}}
    \toprule
    Repo & LOC & Domain & Description & Lang \\
    \midrule
    \href{https://github.com/pymumu/smartdns}{R0} & 1,500k+ & Network & A local DNS server to obtain the fastest website IP. & C/C++ \\
    \href{https://github.com/lss233/kirara-ai}{R1} & 1,000k+ & LLM Agent & A multi-LLM chatbot framework. & Python \\
    \href{https://github.com/elastic/elasticsearch}{R2} & 190,000k+ & Search Engine & A free, open, distributed, RESTful search engine. & Java \\
    \href{https://github.com/soybeanjs/soybean-admin}{R3} & 275k+ & Frontend App & A modern backend admin template. & TS \\
    \bottomrule
  \end{tabular}
\end{table*}

\noindent\textbf{Evaluation Criteria for Personalized Answers.} To comprehensively quantify both the correctness and the degree of personalization in the generated answers, and inspired by prior work on evaluating personalized LLM outputs~\cite{molavi2025llmdrivenpersonalizedanswergeneration, li2024personalizedlanguagemodelingpersonalized}, we established a scoring scheme consisting of five aspects for each question. Each aspect is scored on a 1--10 scale. Correctness ensures the answer is objectively valid (e.g., ``must mention that Class A maintains a buffer''), while the other four aspects—technology stack fit, behavioral habit alignment, understanding level match, and style preference—reflect essential dimensions of how well the answer adapts to the developer’s context. For guidance, we provided participants with a scoring rubric where scores correspond to quality levels (e.g., 0-2: Poor; 3-5: Fair; 6-8: Good; 9-10: Excellent). The detailed scoring criteria, along with the research supporting each aspect, are presented in Table~\ref{tab:scoring-criteria}. 

\noindent\textbf{Evaluation Process For Personalized Answers.} To ensure efficient and accurate scoring against our criteria, we adopted a two-step process.
\begin{enumerate}[leftmargin=2em]
\item We evaluated correctness using an automated scoring approach with an LLM, a method validated in prior studies for assessing answer validity when detailed scoring criteria are provided~\cite{llm-as-judge}. We selected Claude 4 Sonnet for this role due to its strong instruction-following ability and performance on code understanding benchmarks~\cite{anthropic2025claude4report}. Since both the questions and the corresponding repositories are provided by us, we pre-define the key points that should be included in correct answers based on human understanding. Claude 4 Sonnet's task is to compare the human-defined key points with the generated answers and judge whether they matched. All 280 automated results (28 questions × 10 developers) were then manually calibrated for consistency, with 42 scores adjusted via human review to correct errors and resolve ambiguities.
\item To evaluate the four inherently subjective personalization aspects, we first ensured participants had a thorough understanding of the scoring rubric. Then, participants were invited to blindly evaluate the answers generated by baselines and VirtualME enhanced approaches. The comparison experiment solely switched between ``whether the personalized context provided by VirtualME was injected.'' The two sets of generated answers were presented in random order.
\end{enumerate}

\subsection{RQ1 Settings}
The goal of RQ1 is to learn how quickly how quickly and accurately these metrics can converge to reflect developers’ average-level traits. To this end, during the experiment, each of the 10 participants was asked to validate their persona at the end of every week through a structured questionnaire. Specifically, the system presented a set of metrics under each persona dimension. Participants were asked whether system-computed metrics (e.g., "Your average Python productivity: 55 LOC/hour") aligned with their long-term self-perception.

To mitigate biases arising from participants' tendency to overestimate or underestimate themselves, we recruited participants (see Section \ref{subsec:participant-recruit}) with substantial development experience and a more calibrated self-understanding of their behavioral patterns, enhancing the reliability of their feedback.

Based on these responses, we define the accuracy of each persona dimension. We calculate the accuracy for each dimension~$D$ as the proportion of its constituent metrics that a participant confirmed as accurate.
This is formally defined as
\begin{equation}
  \label{eq:accuracy}
  \text{Accuracy}_{D} = \frac{N_{D,\,\text{correct}}}{N_{D,\,\text{total}}},
\end{equation}
where $N_{D,\,\text{correct}}$ is the number of metrics within dimension~$D$ confirmed as correct by the participant, and $N_{D,\,\text{total}}$ is the total number of metrics in that dimension.

\subsection{RQ2 Settings}
The objective of RQ2 is to examine whether the Personalized Repository Q\&A, developed on top of the VirtualME data foundation, can enhance the user’s personalized experience without sacrificing answer correctness. To address this question, we conduct evaluations at the end of Week 4 using the Q\&A dataset constructed in Section 4.2. Our approach is compared against two baseline approaches across two different models.

\textbf{{Comparison Baselines.} }
A suitable baseline must represent the state-of-the-art in repository-level Q\&A, and be widely available and actively maintained, thereby reflecting realistic usage scenarios for practitioners, which are typically integrated into AI IDEs. Based on these criteria, we selected two representative AI-native IDEs:
\begin{itemize}[leftmargin=1em]
    \item \textbf{Cursor (v1.5.11)}~\cite{cursor_v1.5.11}:A VS Code fork with built-in repository indexing and embedded chat. Its Q\&A agent can traverse the AST and consult relevant documentation to answer knowledge-based questions in large codebases.  
    \item \textbf{Trae (v2.3.0)}~\cite{trae_v2.3.0}: Another AI-native IDE, also based on a VS Code fork. Its \texttt{\#workspace} chat mode provides retrieval-augmented answers, serving as an industrial-grade reference for evaluating the benefits of personalization.
\end{itemize}

\textbf{Model Selection.}  
To ensure that any observed improvements are attributable to personalization rather than differences in backbone model capability, the chosen models must be both strong in programming comprehension and widely recognized in prior benchmarks. We therefore adopted the following two state-of-the-art LLMs as shared backbones across all methods:  
\begin{itemize}[leftmargin=1em]
    \item \textbf{Claude 4 Sonnet}~\cite{anthropic2025claude4}: Known for its state-of-the-art performance on programming benchmarks such as HumanEval~\cite{chen2021evaluating} and SWE Bench~\cite{jimenez2023swe}, with strong code comprehension and instruction-following abilities.  
    \item \textbf{GPT 5}~\cite{openai_gpt5}: Representing the latest advancement in code synthesis, repository-level reasoning, and long-context understanding, ranking at the forefront of multiple programming benchmarks.
\end{itemize}

\textbf{Implementation Details.} In this experiment, for consistency, the context window was set to 200k tokens, while other parameters (e.g., temperature, max tokens) followed the default settings of the IDE’s built-in model.

\subsection{RQ3 Settings}
RQ3 investigates whether VirtualME can continuously update developer personas as a developer’s proficiency evolves, and whether such updates enable downstream personalized Q\&A responses to adapt accordingly. To this end, we designed a longitudinal case study with a single participant over a 7-day continuous tracking period. The participant (P9) was a senior Python/TypeScript developer with no prior exposure to Rust. The choice of Rust provided a clear and traceable learning curve, allowing us to observe how the persona evolves alongside skill acquisition. Over 7 days, P9 systematically studied the fundamentals of Rust and engaged in hands-on coding tasks. This setup ensured that the collected behavior data captured realistic exploration, failures, and gradual mastery. On days 1, 4, and 7, we compared the generated answers to the fixed question and repository, analyzing how the content shifted in terms of personalization and technical depth. Post-hoc feedback from the participant was collected to validate whether the observed changes aligned with their own perception of progress.
\begin{itemize}[leftmargin=1em]
    \item \textbf{Repository:} ~\href{https://github.com/gitui-org/gitui}{GITUI}, a fast terminal-ui for git written in rust, which can represent a typical Rust-dominated repository with a certain level of difficulty.
    \item \textbf{Question:} What is the design philosophy behind component traits?
\end{itemize}

\begin{table*}
  \caption{Evaluation Dimensions and Scoring Criteria.}
  \label{tab:scoring-criteria}
  \renewcommand{\arraystretch}{1.0} 
  \small 
  \setlength{\tabcolsep}{3.5pt}   
  \begin{tabular}{p{0.15\linewidth}p{0.7\linewidth}p{0.08\linewidth}}
    \toprule
    \textbf{Aspect} & \textbf{Description} & \textbf{Range} \\
    \toprule
    Correctness~\cite{llm-as-judge} & Each question has predefined scoring points. The score is based on the proportion of points satisfied. & 1--10 \\
    \midrule
    Tech Stack~\cite{qa-benchmark-tech} & Does the answer consider the developer's own technology stack? & 1--10 \\
    \midrule
    Behavioral Habits~\cite{qa-benchmark-beh} & Does the answer consider the developer's behavioral habits? & 1--10 \\
    \midrule
    Understanding Level~\cite{qa-benchmark-und} & Does the answer accurately infer the developer's understanding level of the repository based on their tech stack and known abilities, providing appropriate guidance? & 1--10 \\
    \midrule
    Style \newline Preference~\cite{qa-benchmark-style} & Does the answer accurately infer the developer's preferences and present the information in a suitable style? & 1--10 \\
    \bottomrule
  \end{tabular}
\end{table*}

\tcbset{
  tightbox/.style={
    boxrule=0.5pt, 
    top=4pt, bottom=4pt,  
    left=5pt, right=5pt,     
  }
}

\section{Results}
\label{sec:results}

\subsection{RQ1: Performance on Persona Identificaiton}

\begin{figure*}[t]
  \centering
  \includegraphics[width=0.5\linewidth]{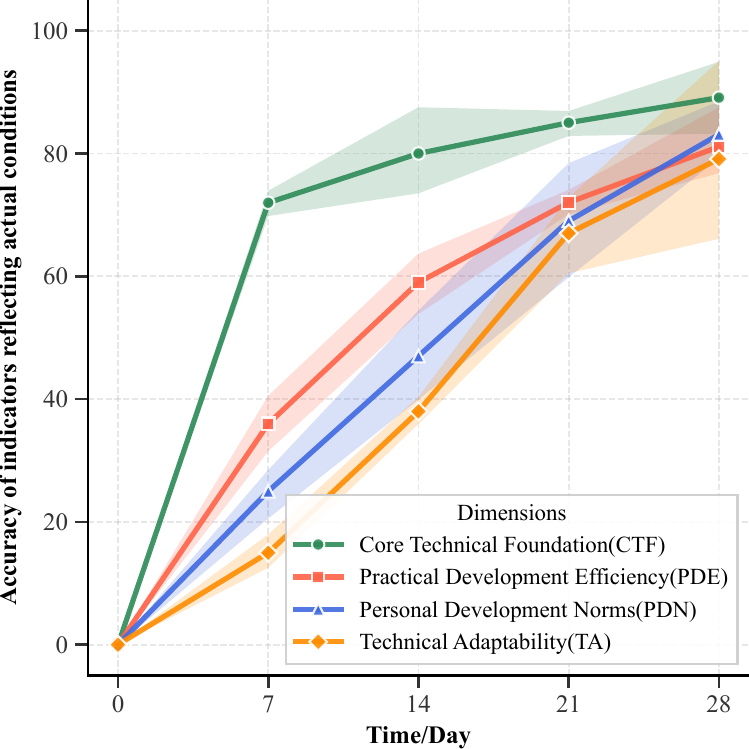}
  \captionof{figure}{Average accuracy of persona indicators in reflecting the developer's actual situation over four weeks.}
  \label{fig:rq1_accuracy}
  
  \vspace{1em} 

  \begin{table}[H]
  \caption{Micro-evolution of P1 Persona (Part).}
  \label{tab:p1-micro}
  \small
  \renewcommand{\arraystretch}{1.0} 
  \begin{tabularx}{\linewidth}{@{} l X c c c @{}}
    \toprule
    \textbf{Day} & \textbf{Frequent Libraries} & \textbf{\begin{tabular}{@{}c@{}}Prod. (LOC/h)\end{tabular}} & \textbf{\begin{tabular}{@{}c@{}}Active Hours\end{tabular}} & \textbf{\begin{tabular}{@{}c@{}}New Libs (lib/wk)\end{tabular}} \\
    \midrule
    0  & —                                & —   & —                      & —               \\
    \midrule
    7  & torchaudio, librosa              & 42  & 14:07--17:58           & 0      \\
    \midrule
    14 & torch, librosa, transformers     & 55  & 09:30--12:00           & 0.6    \\
    \midrule
    21 & torch, librosa, transformers, accelerate\dots & 61  & \begin{tabular}{@{}c@{}}09:30--12:30\\14:05--16:30\end{tabular} & 1.0 \\
    \midrule
    28 & torch, librosa, transformers, accelerate\dots & 60  & \begin{tabular}{@{}c@{}}08:43--12:30\\14:05--18:00\end{tabular} & 1.2 \\
    \bottomrule
  \end{tabularx}
\end{table}
  \label{tab:rq1_eg}

\end{figure*}

After activating VirtualME, all persona metrics exhibited a significant positive convergence trend over time. As shown in Figure~\ref{fig:rq1_accuracy}, the average accuracy across all dimensions reached or exceeded 70\,\% by the third week. Meanwhile, Table~\ref{tab:p1-micro} illustrates the specific evolution of several key metrics within the persona, using the P1 developer as an example.
The metrics in the CTF dimension converged the earliest (approx.~7 days), while those reflecting changes in PDN or TA (e.g., frequency of new technology adoption, active hours) converged the slowest (approx.~28 days).
This is because most metrics in the CTF dimension rely on direct static analysis of code repositories; once a developer opens and performs some activity in these repositories, their external dependencies and programming language composition are immediately used to compute the relevant CTF metrics.
Conversely, metrics that track the frequency of adopting new frameworks, libraries, and technologies, or weekly active hours, require a certain number of data changes to be tracked before a baseline statistical result can be formed.
Therefore, the metrics reflecting a developer's learning and adaptability take longer to stabilize.

\vspace{4pt}

\begin{mdframed}
[linecolor = gray!100,linewidth = 3pt,
innerleftmargin = 3pt, topline=false, rightline=false, bottomline=false, leftline=true, innerrightmargin = 3pt,innertopmargin = 3pt, innerbottommargin = 3pt,backgroundcolor = gray!30]
\textbf{Answering RQ1:} 
With approximately four weeks of data accumulation, VirtualME can reliably fit a developer's core persona attributes (e.g., Knowledge Domain, Active Hours). However, metrics reflecting PDN and TA require a more extended data collection period to achieve stability.
\end{mdframed}

\subsection{RQ2: Performance on Personalized Q\&A }
We investigate whether the personalized Q\&A system built upon the VirtualME data foundation can systematically enhance the user's personalized experience. This inquiry focuses on two core aspects: (1) Is the correctness of the generated answers affected by VirtualME? (2) Does the VirtualME approach significantly improve the personalization level in the answers, making them better aligned with the developer's preferences? Based on the findings from RQ1, we conducted the Q\&A evaluation tasks after the experiment had run for four weeks, including: (1) a correctness evaluation based on an LLM(Claude-4-Sonnet) with manual calibration, and (2) a blind personalization scoring conducted by the participants.

\begin{table*}[h]
  \centering
  \caption{Comparison of Q\&A performance between baseline and VirtualME-enhanced methods. 
  The table shows raw scores for each dimension, the total scores, and the improvement percentage ($\Delta$).}
  \label{tab:rq2_results}
  \small
  \renewcommand{\arraystretch}{1.0}
  \setlength{\aboverulesep}{1pt}
  \setlength{\belowrulesep}{1pt}
  \setlength{\tabcolsep}{1.5pt} 
  \renewcommand{\theadfont}{\bfseries}
  \newcolumntype{C}[1]{>{\centering\arraybackslash}m{#1}}
  
  \begin{tabular}{@{}C{1.3cm} C{2.2cm} *{5}{C{1.15cm}} C{1.2cm} C{1.3cm} @{}}
    \toprule
    \thead{Model} & \thead{Method} & \thead{Corr.} & \thead{Tech\\Stack} & \thead{Behav.\\Habits} & \thead{Und.\\Level} & \thead{Style\\Pref.} & \thead{Total\\Score} & \thead{$\Delta$\\(\%)} \\
    \midrule
    
    % --- Claude-4 Sonnet ---
    \multirow{4}{*}[-0.5ex]{\centering\arraybackslash%
    \parbox{1.2cm}{\centering Claude 4\\[-1pt]Sonnet}}
      & Cursor & 8.48 & 6.01 & 4.96 & 5.86 & 5.54 & 30.85 & -- \\
      & Cursor+VirtualME & \textbf{8.96} & \textbf{8.39} & \textbf{7.96} & \textbf{8.34} & \textbf{8.09} & \textbf{41.74} & \textbf{$\uparrow$35.30} \\
      \cmidrule{2-9}
      & Trae & 8.29 & 5.92 & 5.03 & 5.63 & 5.43 & 30.30 & -- \\
      & Trae+VirtualME & \textbf{8.58} & \textbf{8.58} & \textbf{8.71} & \textbf{8.12} & \textbf{8.54} & \textbf{42.53} & \textbf{$\uparrow$40.36} \\
    \hline
    
    % --- GPT-5 ---
    \multirow{4}{*}[-0.5ex]{\centering GPT 5}
      & Cursor & 8.95 & 6.21 & 5.32 & 5.90 & 5.80 & 32.18 & -- \\
      & Cursor+VirtualME & \textbf{8.96} & \textbf{8.43} & \textbf{8.09} & \textbf{8.29} & \textbf{8.21} & \textbf{41.98} & \textbf{$\uparrow$30.45} \\
      \cmidrule{2-9}
      & Trae & \textbf{8.90} & 5.95 & 5.19 & 5.99 & 6.08 & 32.11 & -- \\
      & Trae+VirtualME & 8.88 & \textbf{8.29} & \textbf{8.18} & \textbf{7.98} & \textbf{8.26} & \textbf{41.59} & \textbf{$\uparrow$29.52} \\
    \hline
    \hline
    
    % --- Summary Rows ---
    \multicolumn{2}{l}{Methods \textbf{Without} VirtualME} & 8.655 & 6.0225 & 5.125 & 5.8450 & 5.7125 & 31.36 & -- \\
    \multicolumn{2}{l}{Methods \textbf{With} VirtualME} & \textbf{8.845} & \textbf{8.4225} & \textbf{8.235} & \textbf{8.1825} & \textbf{8.2750} & \textbf{41.96} & \textbf{$\uparrow$33.80} \\
    
    \bottomrule
  \end{tabular}
\end{table*}

The comparison results between the personalized Q\&A built on the VirtualME data foundation and the baseline methods are shown in Table~\ref{tab:rq2_results}. We can interpret these results from two perspectives: correctness and personalization level. First, regarding correctness, the answers generated with VirtualME did not introduce additional factual errors; their scores were largely on par with the baseline, with some being slightly higher. This indicates that personalization enhancement does not weaken the large model's ability to generate factual content, and the entire injection process maintains correctness without degradation.

Second, in terms of personalization, the advantages of VirtualME were particularly significant. Across the four dimensions—technology stack fit, behavioral habit alignment, understanding level match, and stylistic preference—it achieved marked improvements over the baseline methods, with the overall score improving by an average of 33.80\%. For example, thanks to the comprehensive representation of the developer in the user persona, the agent can determine the user's technology stack from their \texttt{Domain}, assess the fit between the user's stack and the current repository, and infer the match between the user's question and their cognitive state. This allows the agent to modify the standard answer by providing more detailed explanations for difficult concepts or trimming redundant information. This result demonstrates that the introduction of the data foundation can substantially improve the personalized experience of the answers, providing developers with knowledge support that is better suited to their background and needs.

\begin{figure*}[ht]
  \centering
  \begin{subfigure}[b]{0.48\textwidth}
    \centering
    \includegraphics[width=\linewidth]{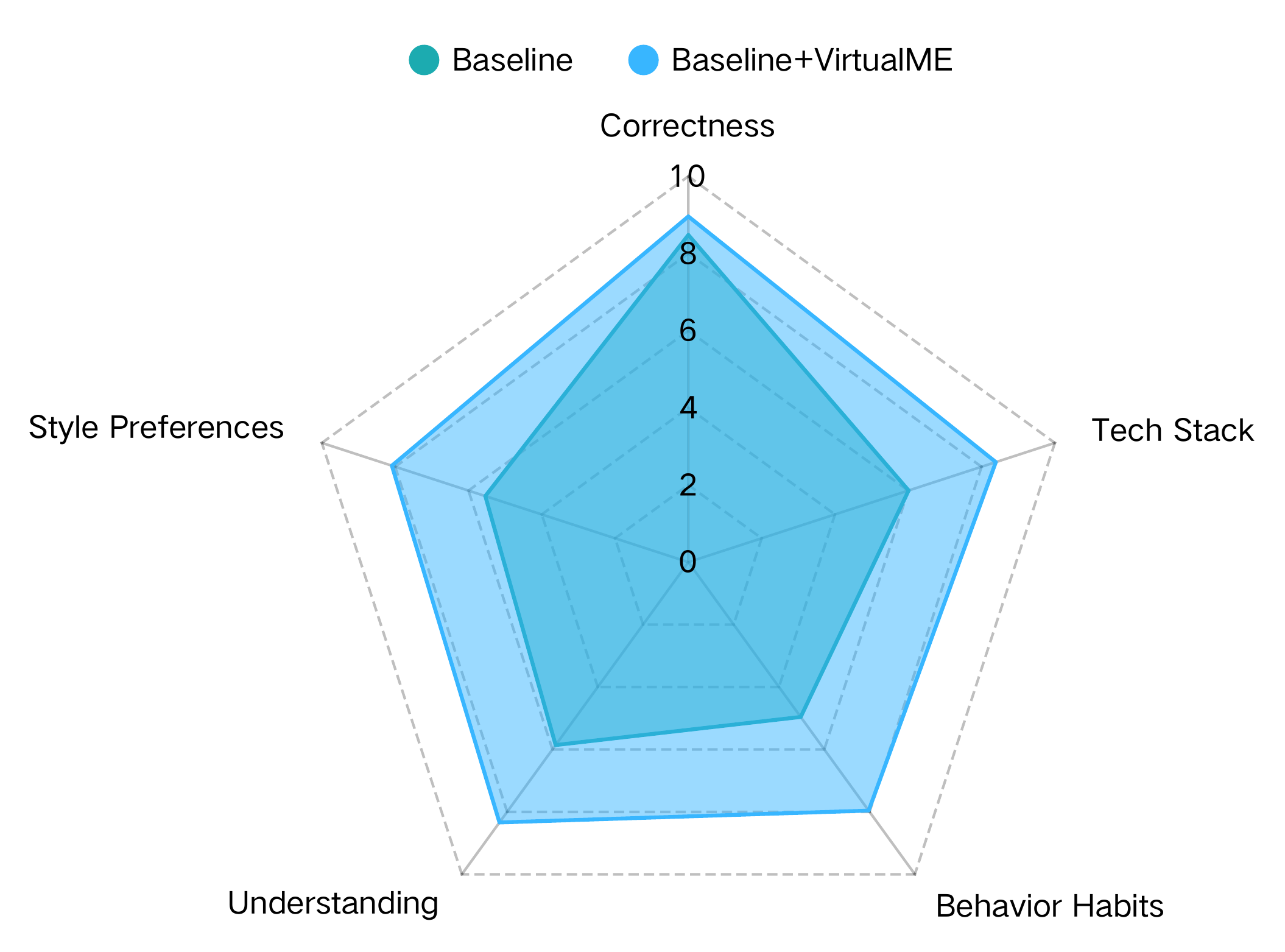}
    \caption{Cursor + Claude 4 Sonnet.}
    \label{fig:radar_cursor_cl4}
  \end{subfigure}\hfill
  \begin{subfigure}[b]{0.48\textwidth}
    \centering
    \includegraphics[width=\linewidth]{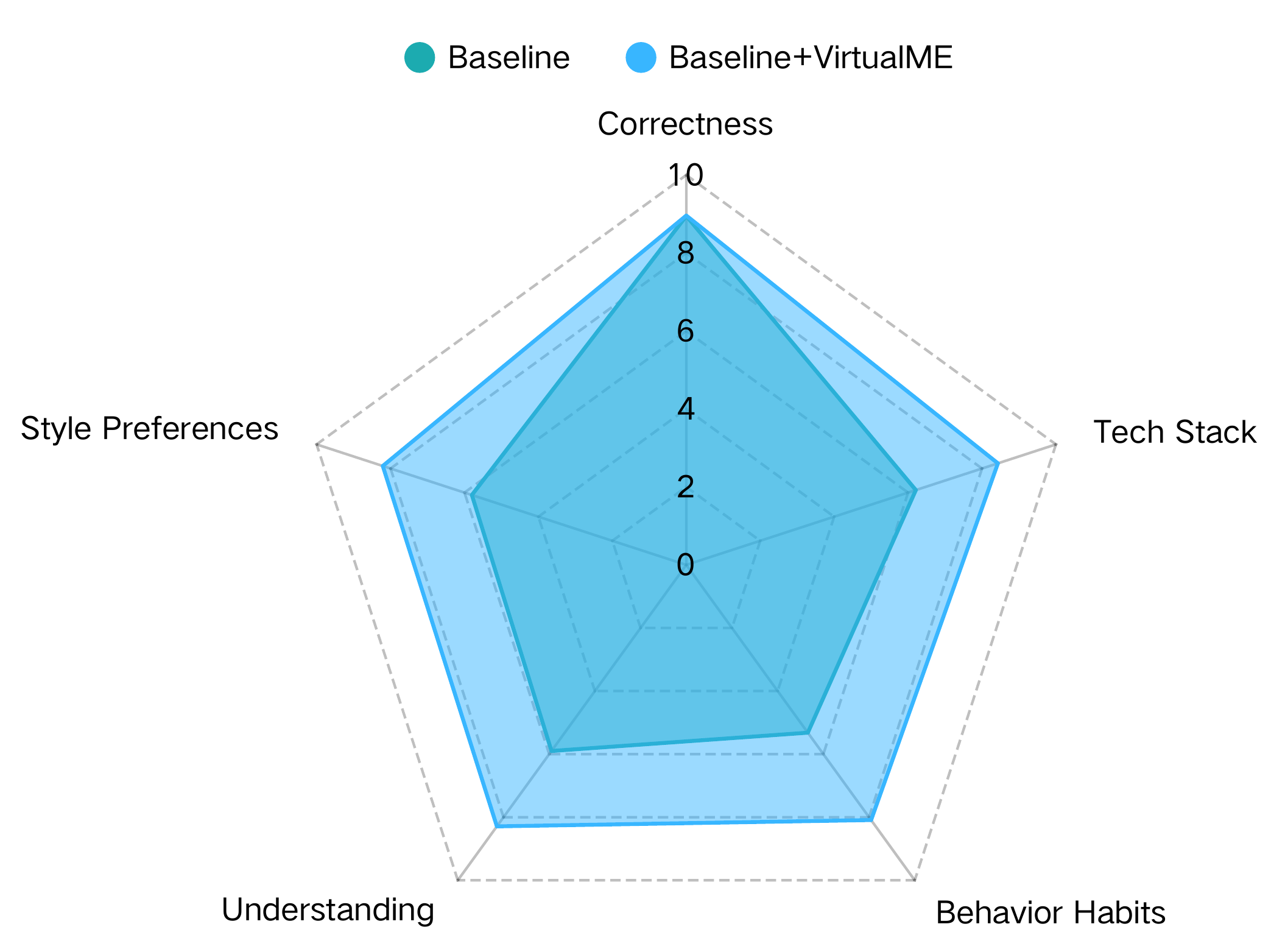}
    \caption{Cursor + GPT 5.}
    \label{fig:radar_cursor_gpt5}
  \end{subfigure}

  \begin{subfigure}[b]{0.48\textwidth}
    \centering
    \includegraphics[width=\linewidth]{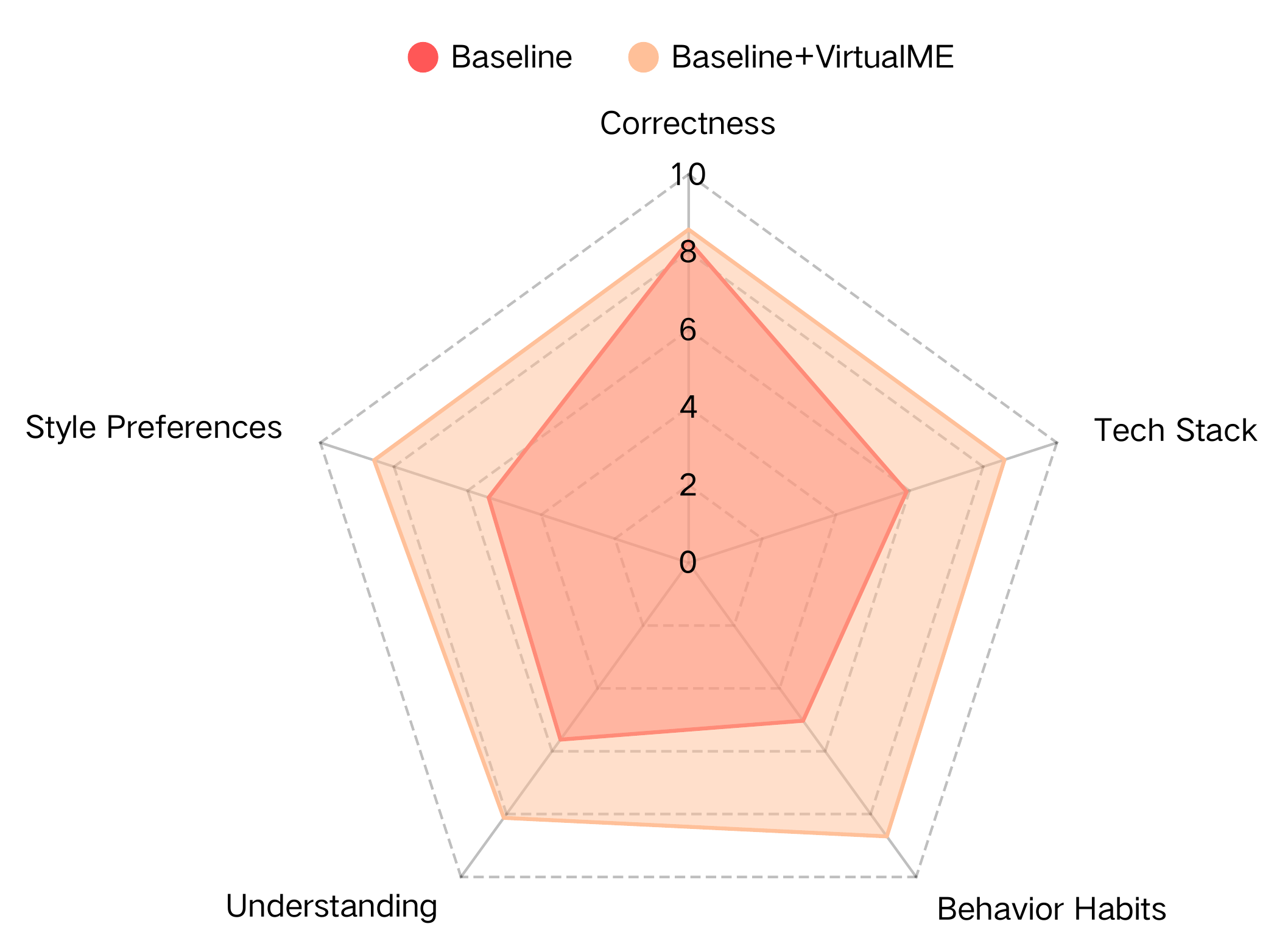}
    \caption{Trae + Claude 4 Sonnet.}
    \label{fig:radar_trae_cl4}
  \end{subfigure}\hfill
  \begin{subfigure}[b]{0.48\textwidth}
    \centering
    \includegraphics[width=\linewidth]{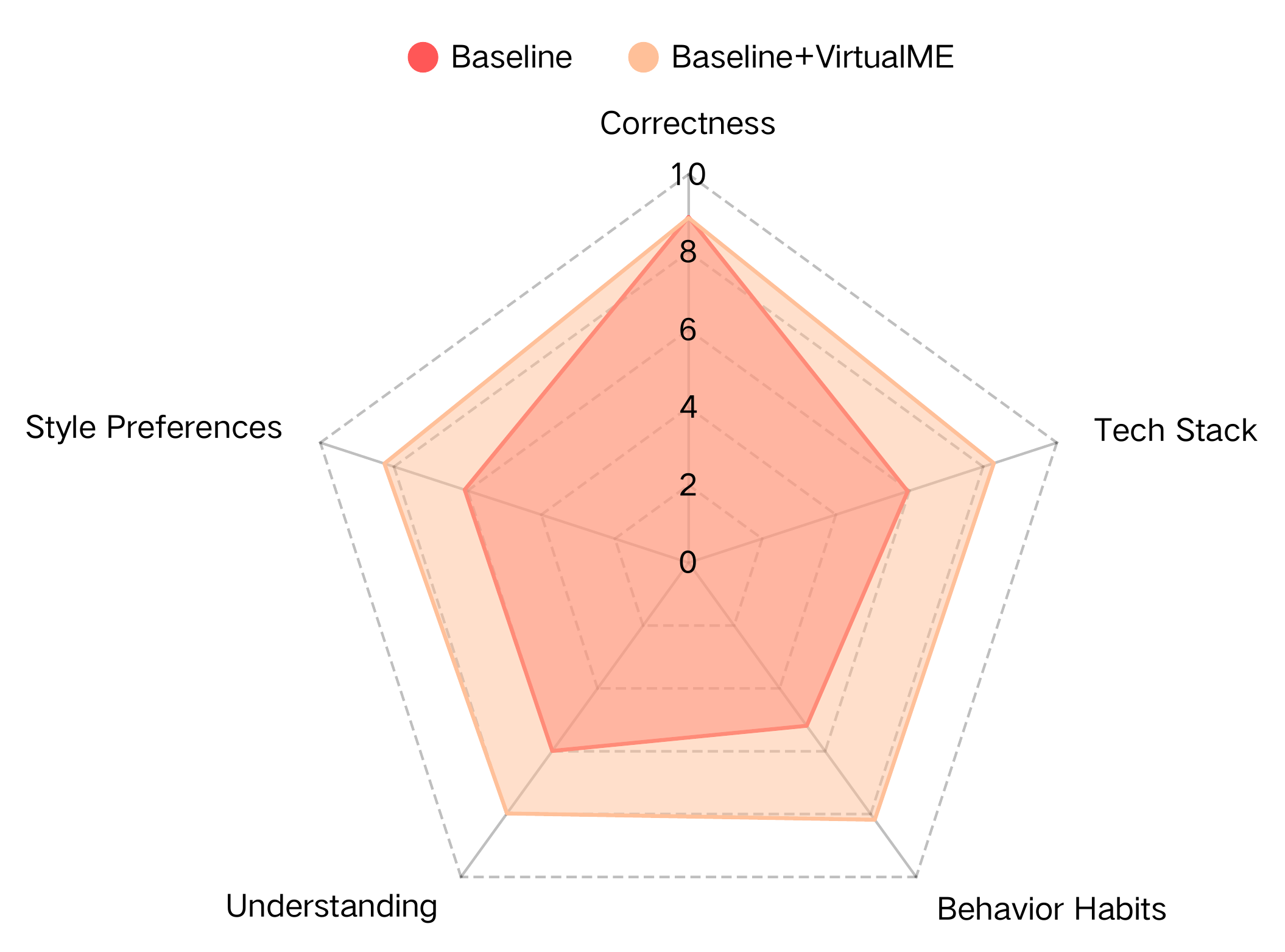}
    \caption{Trae + GPT 5.}
    \label{fig:radar_trae_gpt5}
  \end{subfigure}

  \caption{Repository Q\&A Scores across IDE--LLM pairs. (a) Cursor + Claude 4 Sonnet; (b) Cursor + GPT 5; (c) Trae + Claude 4 Sonnet; (d) Trae + GPT 5.}
  \label{fig:all_radar}
\end{figure*}

\vspace{4pt}
\begin{mdframed}
[linecolor = gray!100,linewidth = 3pt,
innerleftmargin = 3pt, topline=false, rightline=false, bottomline=false, leftline=true, innerrightmargin = 3pt,innertopmargin = 3pt, innerbottommargin = 3pt,backgroundcolor = gray!30]
\textbf{Answering RQ2:} 
In our experimental setup, VirtualME significantly enhanced the personalized experience while ensuring no degradation in factual correctness. All four personalization dimensions saw significant increases, with the total score improving by an average of \textbf{33.80\%}.
\end{mdframed}

\subsection{RQ3: Dynamic Answer Generation for Programmers at Different Skill Levels}
The following presents key fragments and persona updates day-by-day, with conclusions based on text comparison and post-hoc confirmation from the developer P9.

\textbf{Day 1: }
The developer's behavior on the first day was analyzed as follows: the only new Rust library attempted was `std'; Rust-related terminal commands were limited to `cargo new' and `cargo run', with the latter having a 58\% failure rate. The opening paragraph of the agent's answer on that day was:
\vspace{4pt}

\begin{mdquote}\small
``Given your background in Python/TypeScript and your current Rust learning journey, I'll explain this in terms that bridge your existing knowledge…''
\end{mdquote}

Subsequent paragraphs avoided terms like `trait object' and `lifetime', instead using analogies like ``logic'' and ``rendering'', and did not include line-level source code. The developer later confirmed that, at this stage, analogies were indeed necessary to reduce cognitive load, and \textbf{the agent's personalized response allowed them to accept new concepts almost seamlessly}.

\textbf{Day 4: }
The developer's behavior on the fourth day was analyzed as follows: they began to use language features like basic syntax and functional concepts, and the failure rate of Rust-related terminal commands dropped to 18\%. The agent's opening paragraph was revised to:
\vspace{4pt}

\begin{mdquote}\small
``The component system strikes a balance between type safety, performance, and maintainability—key concerns in systems programming with Rust…''
\end{mdquote}
This was followed by the complete `impl DrawableComponent' code block for `CredComponent' (including line numbers), with a comparison to the explicit delegation differences in React. The developer noted that the code snippets and line numbers helped them read the implementation directly, and that analogies were no longer needed at this point.

\textbf{Day 7: }
The developer's behavior on the seventh day was analyzed as follows: Rust file browsing accounted for 71\% of their activity, the terminal failure rate was 9\%, and cyclicality was 0.82; the \texttt{Learning} dimension of the persona was rated 8/10. The agent's answer now began by quoting a design tension from the source code comments:
\vspace{4pt}
\begin{mdquote}\small
``The documentation openly acknowledges a design tension: `It is a little strange that you implement \texttt{draw()} on a \texttt{DrawableComponent}, but have function \texttt{hide()} from trait Component…' ''
\end{mdquote}
\textbf{The rest of the answer contained no more cross-language analogies}, instead directly discussing whether the ``trait separation was over-engineered'' and the possibility of future consolidation. The developer confirmed that they could now understand the original comments and wished to critique the design trade-offs.

\vspace{4pt}

\begin{mdframed}
[linecolor = gray!100,linewidth = 3pt,
innerleftmargin = 3pt, topline=false, rightline=false, bottomline=false, leftline=true, innerrightmargin = 3pt,innertopmargin = 3pt, innerbottommargin = 3pt,backgroundcolor = gray!30]
\textbf{Answering RQ3:} 
When a developer’s proficiency evolves from novice to beginner over the course of a week, the daily-updated persona from VirtualME drove observable changes in the Q\&A responses, indicating that VirtualME can continuously update its user model based on captured developer behavior, enabling downstream agents to evolve in sync with the developer's own characteristics.
\end{mdframed}
\section{Discussion}
\label{sec:discussion}

\subsection{Privacy Concerns}
Our work involves collecting detailed developer behavior data and presenting it to LLM providers. Privacy may be an obstacle to scaling up VirtualME. To protect developers' privacy, we design:
\begin{itemize}[leftmargin=1em]
    \item On-device data processing: All raw IDE interaction logs remain on the user's local machine. The analysis is performed entirely on-device, ensuring that sensitive coding activities never leave the developer's local environment.
    \item Private LLM invocation: When personalization requires LLM-based reasoning, we recommend using enterprise or organization-hosted private LLM rather than public cloud APIs, ensuring that persona data stays within the trusted boundary.
\end{itemize}
 
\subsection{Insights from Empirical Evaluation}
Across 280 paired answers (28 questions × 10 developers), VirtualME improved the median personalization score by 33.80\% while preserving factual correctness, indicating a measurable move toward \emph{Personalized Code Intelligence}.
RQ1's demonstration of high persona fidelity—achieved reliably after four weeks of data accumulation—highlights the advantages of basing personalization on verifiable IDE behaviors.
This fidelity validates the rule-based measurement approach and suggests potential for more robust evaluations of developer characteristics, such as identifying skill gaps or tracking proficiency changes.
Building on this foundation, RQ2 reveals that personalization enhances Q\&A responses by making them more relevant and comprehensible, without undermining their correctness.
This hints at a broader evolution in software engineering, where code intelligence systems should embrace developer diversity to further enhance development efficiency.
RQ3 illustrates VirtualME's adaptability, with developers evolving daily during skill acquisition—from beginner reliance on analogies to assessments of design trade-offs. This temporal responsiveness may support lifelong learning in rapidly evolving domains such as AI frameworks.

\subsection{Future Applications of VirtualME}
Our research demonstrates that the continuous, fine-grained behavioral data generated by developers in the IDE can support the personalization of various downstream AI-assisted coding tasks. Beyond Repository Q\&A, this data holds broad application potential in several other areas:

\begin{enumerate}[leftmargin=1.5em]
    \item \textbf{Bug Fix.} A developer's behaviors before invoking an agent often include the process of reproducing the issue, attempting to solve it, and even revealing the root cause. VirtualME can explicitly provide this latent information to the agent, offering powerful support for bug resolution.
    \item \textbf{Edit Prediction.} A developer's cursor movement logic is correlated not only with code structure but also with behavioral inertia and their own thought process. The fine-grained, real-time observations provided by VirtualME can supply rich data for predicting developer actions or intentions, serving as a training set for relevant predictive models.
    \item \textbf{In-depth Analysis of Developer Behavior.} VirtualME enables a deep dive into the unique behavioral patterns and common mistakes of individual developers. It allows for the observation of their skill maturity evolution, facilitating research focused on the developers themselves.
\end{enumerate}

\subsection{Hyperparameter Settings}
In Section~\ref{subsec:tb}, the use of an agent to analyze a large volume of raw LBs for Task-level Behavior Recognition may introduce performance issues. Summarizing too frequently or analyzing too many LBs at once can lead to unnecessary token consumption, while doing so too infrequently can prevent the accurate recognition of TBs. We found that the timing of the summary operation should be dynamically adjusted based on the developer's activity density. When a developer's actions are sparse over time, more LBs can be processed in a single batch. Conversely, when activity is dense, partitioning by time intervals is more reasonable.

\subsection{Threats to Validity}

To safeguard the rigor of our conclusions, we analyze threats to validity across three key dimensions and outline our mitigations.

\textbf{Construct validity.} While our four-dimensional developer persona offers a structured representation, it may omit scenario-specific traits such as collaborative competence or domain-specific heuristics. To mitigate this, we incorporated weekly self-validation sessions in which developers reviewed and corrected their persona metrics. This iterative feedback loop reduces systematic misrepresentation, and future work may extend the persona with additional dimensions informed by longitudinal studies. On the other hand, developers engaged in different types will impact TA metrics (e.g., “early adopters” receive higher scores than ”maintainers”). This impact would be mitigated if we can observe developers via VirtualME over a long period, since a developer is likely to be assigned different types tasks within a company over time. We plan to extend TA using newly imported libraries and web-browser data in future work.

\textbf{Internal validity.} The stochasticity of the two IDEs (Cursor and Trae) and the two LLMs (Claude-4-Sonnet, GPT-5) may occasionally yield unusually responses independent of personalization, potentially inflating or diluting VirtualME's observed contribution. To mitigate this, we adopted a paired experimental design in which each Q\&A item was answered with and without VirtualME under identical IDE–LLM configurations, and participants performed blind evaluations to eliminate bias. In addition, in scenarios without VirtualME, repeated trials across multiple repositories further averaged out random fluctuations.

\textbf{External validity.} Our 10 recruited developers all had experience with with more than 20k LOC experience, excluding absolute beginners or individuals without formal higher education. Similarly, the benchmark repositories were large, mature projects, omitting legacy codebases, monorepos, or less common languages. To mitigate these constraints, we ensured diversity across domains and programming languages, and our results show consistent improvements across all four repositories. Future replications with novice programmers and more heterogeneous repositories will be necessary to confirm broader applicability. Furthermore, the longitudinal case study in RQ3 involved only a single participant over seven days. To observe more behaviors during one week, we increase their workload burden by assigning a series of progressively challenging tasks, so that VirtualME can capture meaningful shifts and can converge to measure their personalities. We acknowledge that one week limits the generalizability of these findings to broader populations or longer-term skill acquisition processes. We will try to conduct longer, multi-participant longitudinal studies in the future work.

\section{Related Work}
\label{sec:rw}

This section reviews prior research from three complementary perspectives: how developer behavior data has been collected, how such data has been analyzed and modeled, and how repository-level Q\&A has evolved with the rise of LLMs. Together, these strands of work form the background against which our approach is positioned.

\textbf{Developer Behavior Data Collection}. Before the rise of LLMs, technological and methodological constraints meant that studies collecting fine-grained IDE interaction data were relatively scarce~\cite{ioannou2018mining, damevski2016mining, deeb2018using}. Representative tools such as the Enriched Event Stream meta-model~\cite{proksch2018enriched}, FeedBaG~\cite{FeedBaG}, and MIMESIS~\cite{MIMESIS} demonstrated how IDE activities could be recorded along with contextual information, enabling the first wave of data-driven developer studies. With the advent of LLMs in software engineering~\cite{hou2024large, fan2023large}, interest in such in-situ IDE data has been revived~\cite{AthenaLLM, UserMentalModels}.

Inspired by the aforementioned research, our proposed Log-level Behavior Model also adopts a similar structure to collect events and their corresponding contexts. However, compared to prior works, our approach can not only capture basic interaction behaviors of developers within IDEs but also incorporate multi-dimensional information such as project structures and terminal output, thus constructing a more comprehensive behavioral data model.

\textbf{Developer Behavior Analysis and Modeling}. After collecting fine-grained developer behavior data, how to analyze and model it to gain a deeper understanding of developers has become a key focus for researchers. 

Maalej et al.~\cite{maalej2017using} explored data-driven task recommendation, Ciborowska et al.~\cite{ciborowska2020recognizing} applied statistical modeling to infer forthcoming developer activities, and Schmidmaier et al.~\cite{schmidmaier2019real} designed adaptive IDEs that react to real-time behaviors. However, existing work mostly relies on offline datasets for analysis or model training, lacking dynamic utilization of real-time user interaction data, preventing deep personalized recommendations based on developers' real-time behaviors, making it difficult to fully align with individual developers' unique behavioral patterns and real-time needs. In contrast, our approach leverages multi-dimensional data collected from the IDE, and continuously updates high-level behaviors and personalized profiles through LLM agents and rules.

 \textbf{Repository Q\&A}. Parallel to these efforts, repository Q\&A has become a key research frontier. Recent benchmarks such as CodeRAG-Bench~\cite{wang2024coderag} and CoReQA~\cite{chen2025coreqa} emphasize the need for cross-file reasoning, while systems like RepoUnderstander~\cite{ma2025alibabalingmaagentimprovingautomated} constructs repository knowledge graphs and employs a Monte Carlo Tree Search-based strategy to achieve holistic understanding of code repositories. These approaches all share a common focus: enhancing the agent's understanding of the \emph{codebase}. In contrast, our work is founded on the principle that in repository-level Q\&A, the most critical context is not just the code, but the \emph{developer}. Our approach is the first to systematically model a developer's dynamic, in-IDE behaviors to construct a multi-faceted persona, and then inject this persona into the Q\&A process. This shifts the paradigm from being merely repository-aware to being truly developer-centric.

\section{Conclusion}
\label{sec:conclusion}

This work highlights a critical limitation in code intelligence research: the neglect of individual developer differences. We argue that effective assistance requires moving beyond one-size-fits-all solutions toward adapting to developers’ unique characteristics. To this end, we proposed \textbf{VirtualME}, an IDE-embedded infrastructure that transforms fine-grained IDE traces into task-level behaviors and comprehensive developer personas. 
Building on this foundation, we developed a repository-level Q\&A module that adapts to developers’ expertise, preferences, and habits. Our evaluation on four repositories and four weeks of traces from ten developers shows that VirtualME improves answer quality and personalization, with an average 33.80\% gain over baselines. These results demonstrate the potential of IDE behavioral data as a foundation for \textbf{Personalized Code Intelligence}.
While promising, this work represents only the first step. Future research will focus on: (1) more intelligent context retrieval that integrates with agent reasoning, (2) generalizing personalization to tasks such as code generation and review, and (3) deeper observation of developer habits and growth to support self-assessment. We release the \textsc{VirtualME-Trace} dataset and benchmarks to foster reproducible research.

\section{Data Availability}
\label{sec:data}
The dataset of this paper is publicly available at \href{https://virtualme26.github.io/VirtualME/}{https://virtualme26.github.io/VirtualME}.
\section*{Acknowledgments}
\label{sec:ack}
This work is supported by Huawei, the National Natural Science Foundation of China (Grant Nos. 62332001, 62272445) and the Software Modeling Project under Grant ZC25T320070-31. We thank Hao Wang, Bangxi Yin and Hongjun Yang for their contributions to the demo development.

\bibliography{ref}
\bibliographystyle{ACM-Reference-Format}

\end{document}